\documentclass [aps,pra,twocolumn,amsfonts]{revtex4}
\usepackage{graphicx}

\def\BEA {\begin{eqnarray}}
\def\EEA {\end{eqnarray}}
\def\BE {\begin{equation}}
\def\EE {\end{equation}}
\def\BA {\begin{array}}
\def\EA {\end{array}}
\def\NN {\nonumber}
\def\ep{\varepsilon}
\begin{document}

\title{Entanglement-enhanced classical capacity of two-qubit quantum channels with memory: the exact solution}

\author{D.~Daems}

\affiliation{Quantum Information and Communication, Ecole Polytechnique, CP 165, Universit\'e Libre de Bruxelles, 1050 Brussels, Belgium}

\begin{abstract}
The maximal amount of  information which is reliably transmitted over two uses of general  Pauli channels with memory is proven to be achieved by maximally entangled states beyond some memory threshold.
In particular, this proves a conjecture on the depolarizing channel by Macchiavello and  Palma [Phys. Rev. A {\bf 65}, 050301(R) (2002)].
Below the memory threshold, for arbitrary Pauli channels, the two-use classical capacity is only achieved by a particular type of product states.
 \end{abstract}

\maketitle\texttt{}
The transmission of information over long distances in devices like optical fibers or the storage of information in some type of memory are tasks of quantum information processing that can be described by quantum channels.
A major problem in quantum information theory is the evaluation
 of the classical capacity of quantum communication channels, which represents the amount of classical information which can
be reliably transmitted by quantum states in the presence of a noisy environment.  
Early works in this direction  were mainly devoted to memoryless channels for which consecutive signal
transmissions through the channel are not correlated
\cite{H73}-\cite{A05}.
Recently, much attention was given to quantum channels
with memory \cite{Macch1}-\cite{KDC06} in the hope that by entangling multiple uses of the channel, a larger amount of classical information per use could be reliably transmitted. 
For   bosonic continuous
variable memory channels,   entangled states are shown  to enhance the channel capacity  \cite{GM05}-\cite{RSGM05} except in the absence of input energy
constraints.
Moreover, when the memory is modelled as a correlated noise, for each value
of the noise correlation parameter, there exists an optimal degree
of entanglement  that maximizes the
channel capacity \cite{CCMR05}. 
For qubit channels with
memory  it was shown that
maximally entangled states enhance the
 two-use channel capacity with respect to product states if the correlation is stronger than some critical value.
This was conjectured for the depolarizing channel with memory \cite{Macch1} and proven for a particular Pauli channel \cite{Macch2}.
An intriguing open question we shall address is whether for some Pauli channels with memory the capacity could be achieved by progressively entangling two uses of the channel, as occurs for some Gaussian channels where no threshold of correlations is present. 
We prove here that the  states which optimize the transmission of classical information over two uses of any Pauli channel with memory modelled as a correlated noise are particular product states below some memory threshold, and maximally entangled states above that threshold.

The action of $n$ uses of a transmission channel on an initial state $\rho$
is described by a completely positive map ${\mathcal E}$ which can be represented as an operator-sum
\begin{equation}
  \rho \rightarrow {\mathcal E}(\rho) = \sum_k A_k\rho A_k^\dag,
                     \qquad \sum_k A_k^\dag A_k= \rm{id} .
\end{equation}
The amount of classical information which is reliably transmitted
by quantum states through the channel is given by the Holevo-Schumacher-Westmoreland
bound \cite{H73} 
\begin{equation}\label{HSW}
   \chi(\mathcal E) = \max_{\{p_i,\rho_i\}}  \left( S\left(\sum_i p_i\mathcal E(\rho_i)\right)
                     -\sum_i p_i S(\mathcal E(\rho_i))
                \right),
\end{equation}
where $S(\rho) =- {\rm Tr} (\rho \log_2\rho)$ is
the von Neumann entropy and the  maximum is taken over all  ensembles
of input states $\rho_i$ with {\it a priori} probabilities
$p_i$.
The $n$-use classical capacity of the channel is this amount of reliably transmitted information per use 
\BE
{\mathcal C}_n({\mathcal E})= \frac{1}{n} \chi({\mathcal E}), \label{Cn}
\EE
whereas the classical capacity  is defined as ${\mathcal C}=\sup_n {\mathcal C}_n$.
Here we focus on the case of two uses of a single qubit channel with memory considered in Refs. \cite{Macch1},\cite{Macch2}
\BE
{\mathcal E}(\rho)= \sum_{i,j=0}^3p_{ij}\,\sigma_i\otimes\sigma_j \ \rho \ \sigma_i\otimes\sigma_j , \label{Erho}
\EE
where $\sigma_0$ denotes the identity and \{$\sigma_1,\sigma_2,\sigma_3\}$  are the Pauli matrices.
The memory is modelled as a correlated noise such that with probability $\mu \in [0,1]$ the same random Pauli transformation is applied to both qubits while with probability $1-\mu$ the two rotations are uncorrelated
\BE
\label{p}
  p_{ij}=(1-\mu)\,q_i q_j+\mu\,q_i \delta_{ij}, \quad \sum_{j=0}^3q_{j}=1. 
\EE
As the maximally mixed state $\frac{1}{4}\openone$ gives the largest possible entropy $S(\frac{1}{4}\openone)=\log_2(4)$, from the definitions (\ref{HSW}) and (\ref{Cn}), the 2-use classical capacity is upper bounded by \cite{Macch2}
\begin{equation}
\label{C2}
{\mathcal C}_2({\mathcal E})
 \leq 1- \frac{1}{2} S({\mathcal E}(\rho_*)),
\end{equation}
where $\rho_*$ denotes an input state which minimizes the output entropy when transmitted through the channel ${\mathcal E}$.
This upper bound (\ref{C2}) can be achieved in any channel  whose action consists of random tensor products of Pauli transformations such as (\ref{Erho}) provided an input state $\rho_*$  can be identified \cite{Macch2}.
In a nutshell, the argument  amounts to constructing from $\rho_*$ an  ensemble of input states $\sigma_i\otimes\sigma_j  \rho_\star  \sigma_i\otimes\sigma_j$ which each have the same output entropy. 
On the other hand, for such an ensemble taken with equal \emph{a priori} probabilities, one can show that the output state is maximally mixed. 
To optimize the  transmission of information in Pauli channels with memory all that is required is thus to identify an \emph{optimal} input state $\rho_*$. 
Moreover, by the concavity of the Von Neuman entropy, this search can be restricted to pure input states $\rho_*=|\Psi_\star \rangle \langle \Psi_\star | \equiv \rho_{\Psi_\star}$ \cite{Macch2}.

To date, the optimality of some input states has been conjectured \cite{Macch1} for the depolarizing channel ($q_0\!=\!1-p, q_1\!=\!q_2\!=\!q_3\!=\!p/3$) and  proven \cite{Macch2} only in one particular instance of Pauli channel with memory ($q_0\!=\!q_3\!=\!p, q_1\!=\!q_2\!=\!\frac{1}{2}-p$). To study the nature of the optimal states for arbitrary Pauli channels, we consider the  two-qubit pure state obtained from the general superposition
\BE
|\Psi \rangle=c_{00} |00\rangle+c_{11} e^{i \varphi_{11}} |11 \rangle+c_{10} e^{i \varphi_{10}} |10 \rangle+c_{01} e^{i \varphi_{01}} |01 \rangle. \label{psi}
\EE 
The normalization implies the relation $c_{00}^2+c_{11}^2+c_{10}^2+c_{01}^2=1$.
This constraint is taken into account here by expressing the pertaining parameters in terms of three angles $\theta$, $\phi$ and $\psi$ as follows
\BEA
c_{00}&=&\cos \frac{\phi+\psi}{2} \cos \frac{\theta}{2} \NN\\
c_{11}&=&\sin \frac{\phi-\psi}{2} \sin \frac{\theta}{2} \NN\\
c_{10}&=&\cos \frac{\phi-\psi}{2} \sin \frac{\theta}{2} \NN\\
c_{01}&=&\sin \frac{\phi+\psi}{2} \cos \frac{\theta}{2}.
\label{tfp}
\EEA
The density matrix $\rho_{\Psi}$ can be expressed in terms of the tensor products of Pauli matrices  as  
\BE
\rho_{\Psi}=\frac{1}{4}\sum_{n,k=0}^3 w_{nk} \sigma_n\otimes\sigma_k, \label{dec}
\EE
with the real coefficients $w_{nk}$ given by
\BE
w_{nk}=\rm{Tr} (\rho_{\Psi} \ \sigma_n \otimes \sigma_k).
\label{wnk}
\EE
Note that $w_{00}=\rm{Tr} (\rho_{\Psi})=1$ and, by the Schwartz inequality, $|w_{nk}| \leq 1$.
The purity $\Pi(\rho_{\Psi})\equiv{\rm Tr}(\rho^2_{\Psi})$ is unity for a pure state  which is accounted for by the relation
\BE
\sum_{k=1}^3 w_{kk}^2+\sum_{n \neq k=0}^3 w_{nk}^2=3. \label{sumw}
\EE 
In addition, the following inequality holds for any permutation of the indexes 1, 2, 3
\BE
\label{ineq}
w_{jj}^2+w_{kk}^2- w_{nn}^2 \leq 1 . 
\EE
For instance, from the explicit expression of (\ref{wnk}) for each $w_{kk}$ in terms of the angles and phases entering (\ref{psi})-(\ref{tfp}), one arrives at
\begin{widetext}
\BEA
\alpha_{\pm}&\equiv& w_{11}^2+w_{22}^2\pm w_{33}^2\NN\\
&=&\frac{1}{2} \left\{ \cos^2(\varphi_{10}-\varphi_{01}) [\sin \phi +\sin \psi]^2 + \cos^2 \varphi_{11} [\sin \phi -\sin \psi]^2 \right\} \pm (\cos \theta \cos \phi \cos \psi-\sin \phi \sin \psi)^2.
\label{alp}
\EEA
\end{widetext}
This yields the tight bounds $\alpha_-\leq 1$, which proves one version of (\ref{ineq}), and $\alpha_+ \leq 3$.
Another  quantity that will be relevant below is
\BEA
\beta&\equiv& \sum_{n=1}^3 \left(w_{n0}^2+w_{0n}^2\right)
\NN\\
&=&2\left(1-\sin^2 \theta [\sin^2 \psi \cos^2 \varphi+\sin^2 \phi \sin^2 \varphi ] \right) , \
\label{bet}
\EEA
with the notation $\varphi\equiv(\varphi_{10}+\varphi_{01}-\varphi_{11})/2$.
Clearly, the bound $\beta\leq 2$ is tight.

The interest of the decomposition (\ref{dec}) is that, in addition to the channel-independent relations (\ref{sumw})-(\ref{bet}),
the action of the channel takes on the simple form
\BE
{\mathcal E}(\sigma_n\otimes\sigma_k) =\ep_{nk} \ \sigma_n\otimes\sigma_k, \label{Enk}
\EE
with the \emph{channel parameter} $\ep_{nk}\equiv \sum_{i,j=0}^3p_{ij}\,s_{in} s_{jk} \in [-1,1]$.
This stems from the identity $\sigma_j \sigma_k \sigma_j= s_{jk}\, \sigma_k$
 where $s_{nk}=+1$ if either $n=k$ or $n=0$ or $k=0$, and $s_{nk}=-1$ otherwise.
With  the joint probability (\ref{p}),
the channel parameters read
\BE
\ep_{kk^\prime}=(1-\mu)\,\ep_k \ep_{k^\prime}+\mu\,\ep_{k^{\prime\prime}}, \label{epnk}
\EE
where $k^{\prime\prime}$ is the index of the matrix  $\sigma_{k^{\prime\prime}}$ to which $\sigma_k \sigma_{k^{\prime}}$ is proportional ({\it i. e.}, $k^{\prime\prime}=0$ if $k=k^{\prime}$, $k^{\prime\prime}=k^\prime$ if $k=0$, $k^{\prime\prime}=k$ if $k^\prime=0$ and $k^{\prime\prime}=\{1,2,3\} \backslash  \{k,k^\prime\}$ otherwise). Notice that $\ep_{nk}=\ep_{kn}$. We define the channel parameter
\BE
\ep_n=\sum_{k=0}^3 q_k s_{kn},\label{epn}
\EE
which implies  that $\ep_0=\ep_{00}=1$ and $\ep_{k0}=\ep_k$. 
The ordering of the channel parameters (\ref{epnk})-(\ref{epn}) will turn out to be crucial.
For that purpose, we introduce the non-zero indexes  $l$ (large), $m$ (medium) and $s$ (small) by
\BE 
|\ep_l | \geq |\ep_m| \geq |\ep_s |. \label{sml}
\EE 
The following properties then hold for any value of $\mu$
\BEA
\ep_{ll}^2 &\geq& \ep_{kk^\prime}^2  \qquad \forall k, k^{\prime}\neq 0 \label{epll}\\
\ep_l ^2 &\geq& \ep_{kk^\prime}^2 \qquad   \ \forall k\neq k^{\prime} \label{epl},
\EEA
since $ \ep_k  \ep_{k^\prime} \leq  \ep_l^2  \leq |\ep_l| $ for all $ k, k^{\prime}\neq 0$ and  $\ep_{k^{\prime\prime}} \leq |\ep_l|  \leq 1$ for all $k\neq k^\prime$ (as $\ep_{k^{\prime\prime}}=1$ otherwise).

In order to identify the states $\rho_{\Psi_\star}$  whose output entropy $S({\mathcal E}(\rho_{\Psi_\star}))$  is minimal, the eigenvalues of ${\mathcal E}(\rho_{\Psi})$ are to be considered.  
In terms of the decomposition (\ref{dec}) and of the mapping (\ref{Enk}),
the  channel (\ref{Erho}) reads explicitly ${\mathcal E}(\rho_{\Psi})=\sum_{f s f^\prime s^\prime=0,1} |f \, s\rangle \langle f \, s|{\mathcal E}(\rho_{\Psi})|f^\prime \, s^\prime\rangle \langle f^\prime \, s^\prime|$ with 
\begin{widetext}
\BEA
\langle f \, s|{\mathcal E}(\rho_{\Psi})|f \, s\rangle &=&\frac{1}{4} +(-1)^{f}\ep_{3}w_{30} +(-1)^{s}\ep_{3}w_{03}+(-1)^{f+s}\ep_{33}w_{33}\NN\\
\langle f \, s|{\mathcal E}(\rho_{\Psi})|f+1 \, s \rangle &=&\ep_1 w_{10}-i(-1)^f \ep_{2}w_{20} +(-1)^{s}\ep_{13}w_{13} -i(-1)^{f+s} \ep_{23}w_{23} \NN\\
\langle f \, s|{\mathcal E}(\rho_{\Psi})|f \, s+1 \rangle &=& \ep_{1}w_{01} -i(-1)^s \ep
_2 w_{02}+(-1)^{f}\ep_{13}w_{31} -i(-1)^{f+s} \ep_{23}w_{32} \NN\\
\langle f\, s|{\mathcal E}(\rho_{\Psi})|f+1 \, s+1\rangle &=& \ep_{11}w_{11}-i(-1)^{f}\ep_{21}w_{21}  +i(-1)^{s}\ep_{12}w_{12} -(-1)^{s+f}\ep_{22}w_{22}.
\EEA
\end{widetext}
The roots of the pertaining characteristic equation  $\lambda^4-\lambda^3+a_2\lambda^2+a_1\lambda^1+a_0=0$ are given by \cite{AS}
\BEA
\lambda_{\eta,  \upsilon}=\frac{1}{4}(1 + \eta R +\upsilon Q_{\eta}) \qquad \eta, \upsilon=\pm 1 ,\
\EEA
where  $R=\sqrt{1-4 a_2+ \omega(\{a_i\})}$ and $Q_{\eta}=\sqrt{2-4a_2+\eta(4a_2-8a_1-1)/R-\omega(\{a_i\})}$.
The function $\omega$ need not be specified here.
Owing to the symmetric  structure of the roots in $\eta$, $\upsilon=\pm 1$, it can be shown that extrema in the eigenvalues can be achieved if and only if the coefficients $a_i$ are each extremal.
To minimize the output entropy, the quantities $R$ and $Q_{\eta}$ have to be maximized. 
As $a_2$ is positive  and enters both $R$ and $Q_{\eta}$ with a negative sign, this coefficient has to be minimized.
It reads $a_2=\frac{1}{8}(3-A-B-C)$ with
\BEA
A\!&\!=\!&\!\ep_{ll}^2w_{ll}^2+\ep_{mm}^2w_{mm}^2+\ep_{ss}^2w_{ss}^2\\
B\!&\!=\!&\!\ep_l^2(w_{0l}^2+w_{l0}^2)+\ep_m^2(w_{0m}^2+w_{m0}^2)+\ep_s^2(w_{0s}^2+w_{s0}^2)\NN\\
C\!&\!=\!&\!\ep_{lm}^2(w_{lm}^2+w_{ml}^2)+\ep_{ls}^2(w_{ls}^2+w_{sl}^2)+\ep_{ms}^2(w_{ms}^2+w_{sm}^2). \NN
\label{ABC}
\EEA
The optimal states are those that maximize $A+B+C$.
Their identification rests on two elements:  the constraints imposed on the \emph{weights} $w_{nk}^2$ associated with the decomposition (\ref{dec}) and, on the other hand, the ordering of the channel parameters $\ep_{kk}$ and $|\ep_{l}|$ which is not covered by (\ref{epll})-(\ref{epl}) and depends on $\mu$ through (\ref{epnk}).
Firstly, the sum of all  the weights $w_{nk}^2$ involved in $A+B+C$ is equal to 3 by the pure state identity (\ref{sumw}).
The weights featured in $A$  sum to $\alpha_+ $ by its definition (\ref{alp}), those entering $B$ sum to $\beta$  by (\ref{bet}) and those of $C$ sum thus to $3-\alpha_+-\beta$.
From the parametrization (\ref{psi})-(\ref{tfp}) of the pure state $\rho_{\Psi}$, we derived the bounds $\alpha_+ \leq 3$ and $\beta\leq 2$.
These bounds are tight but  cannot be achieved by the same pure state (since $\alpha_+ + \beta \leq 3$).
They imply that the 3 weights involved in $A$ can be saturated for some states whereas in  $B$  at most 2 of the 6 weights can be equal to unity. 
Note that the positivity of $a_2$ also stems from the purity identity (\ref{sumw}) together with the property $\ep_{nk}^2 \leq 1$.

Secondly, the degree of correlation modifies the  positions of the channel parameters $\ep_{kk}$ with respect to  $|\ep_{l}|$. When $\mu$ goes from 0 to 1, each $\ep_{kk}=(1-\mu) \ep_k^2+\mu$ increases from $\ep_k^2$ to 1. Since $\ep_k^2 \leq |\ep_k| \leq 1$  and, by definition, $|\ep_k| \leq |\ep_l| $, there are values of $\mu$ where $\ep_{kk}$ crosses $|\ep_{l}|$. The ordering (\ref{sml}) also entails that $\ep_{ll}\geq \ep_{mm}\geq \ep_{ss}$ for any $\mu$.

On combining these two aspects we are led to distinguish several intervals of the memory parameter.
For $0 \leq \mu \leq \mu_{ml}$ one has $\ep_l^2 \geq \ep_{mm}^2 $.
Applying the inequality (\ref{epll}) to the definitions (\ref{ABC}) yields $A+C \leq \ep_{ll}^2 (3-\beta)$. Similarly, (\ref{epl}) leads to $B \leq \ep_l^2 \beta$.
Hence, we obtain
\BEA
A+B+C &\leq& 3 \ep_{ll}^2  + \beta (\ep_{l}^2 -\ep_{ll}^2 )\NN\\
&\leq& 3 \ep_{ll}^2  + \beta (\ep_{l}^2 -\ep_{mm}^2 )\NN\\
&\leq&  \ep_{ll}^2 +2 \ep_l^2.
\label{ABC1}
\EEA
On the second line use was made of the relation $\ep_{mm}^2 \leq \ep_{ll}^2$ to introduce the factor $\ep_{l}^2 -\ep_{mm}^2$ which is positive in this region of $\mu$. Hence the corresponding term is majorized by taking the upper bound $\beta=2$.
The bound (\ref{ABC1}) is tight and achieved if and only if  $w_{ll}^2=w_{0l}^2=w_{l0}^2=1$ which characterizes the optimal states.
 The threshold   $\mu_{ml}\equiv (|\ep_l|-\ep_{m}^2)/(1-\ep_m^2)$ is the value of $\mu$ such that $\ep_{mm}^2=\ep_l^2$.
 
This result can be understood as follows.
In this interval of the memory parameter, $\ep_l^2$ and $\ep_{ll}^2$ are larger than any other  $\ep_{kn}^2$.
These channel parameters are associated with precisely three weights:  $w_{0l}^2$ and $w_{l0}^2$ which are featured in $B$ and $w_{ll}^2$ which is featured in $A$.
The optimum is thus $A+B+C=2\ep_{l}^2+\ep_{ll}^2$ and it is reached only for $w_{ll}^2=w_{0l}^2=w_{l0}^2=1$.
Recalling (\ref{dec}), the optimal states are product states of the form
\BEA
\rho_{\Psi_\star}&=&\frac{1}{4} \left( \sigma_0+\zeta \, \sigma_l \right)\otimes \left( \sigma_0+\xi \, \sigma_l \right) \qquad  \zeta,\xi=\pm 1, \ \NN\\
&=&|\Psi_{l,\zeta}\rangle \langle \Psi_{l,\zeta}| \otimes |\Psi_{l,\xi}\rangle \langle \Psi_{l,\xi}| , \
\label{optimpro}
\EEA 
where $|\Psi_{l,\xi}\rangle$ is a single qubit  eigenstate of $\sigma_l$, \emph{i.e.}, $\sigma_l |\Psi_{l,\xi}\rangle =\xi |\Psi_{l,\xi}\rangle$.
For low correlations  the optimal states are therefore not any product states but those which correspond to the eigenstates associated with the  channel parameter $\ep_l$ of largest absolute value. 
The eigenvalues of ${\mathcal E}(\rho_{\Psi_\star})$, required to calculate ${\mathcal C}_n({\mathcal E})$ from (\ref{C2}), are
\BEA
\lambda_{\eta,  \upsilon}=\frac{1}{4}(1 + \eta \ep_{ll} +\upsilon [1+\eta]\ep_l) \qquad \eta, \upsilon=\pm 1 .\
\label{valpro}
\EEA

In the interval   $\mu_{ml} \leq \mu  \leq \mu_\star$, the  ordering with respect to $\ep_l$ is  $\frac{1}{2}(\ep_{ss}^2+\ep_{mm}^2) \leq \ep_l^2 \leq \ep_{mm}^2$.
From (\ref{ABC}) and (\ref{epl}) we obtain $B+C \leq  \ep_l^2(3-\alpha_+)$, and, subsequently
\begin{widetext}
\BEA
A+B+C &\leq&  \ep_{ll}^2+2 \ep_l^2 +(\ep_l^2-\ep_{ll}^2)(1-w_{ll}^2)+(\ep_l^2-\ep_{mm}^2)(w_{ss}^2-w_{mm}^2)+w_{ss}^2(\ep_{ss}^2+\ep_{mm}^2-2 \ep_{l}^2) \NN\\
&\leq&  \ep_{ll}^2+2 \ep_l^2 +(\ep_l^2-\ep_{mm}^2)(1-w_{ll}^2-w_{mm}^2+w_{ss}^2)+w_{ss}^2(\ep_{ss}^2+\ep_{mm}^2-2 \ep_{l}^2)\NN\\
&\leq&  \ep_{ll}^2+2 \ep_l^2 .
\label{ABC3}
\EEA
\end{widetext}
The second inequality rests on the facts that $w_{11}^2 \leq 1$ and $\ep_{mm}^2 \leq \ep_{ll}^2$.
The final result arises because  the factor $1-w_{ll}^2-w_{mm}^2+w_{ss}^2$ is positive or zero by (\ref{ineq}) while  $\ep_l^2-\ep_{mm}^2$ is negative in this region of $\mu$, and similarly for $\ep_{ss}^2+\ep_{mm}^2-2 \ep_{l}^2$ which is negative.
The  tight bound (\ref{ABC3}) coincides with (\ref{ABC1})  and is realized  iff $w_{0l}^2=w_{l0}^2=w_{ll}^2=1$, \emph{i.e.}, for the optimal states (\ref{optimpro}).

Notice that, in contrast to the previous interval,   here $\ep_{mm}^2 \geq \ep_l^2$.
Hence, one might have been tempted to consider a state which saturates $w_{mm}^2$ instead of both $w_{0l}^2$ and $w_{l0}^2$ whose prefactor in $B$ is $\ep_l^2$.  However, a state characterized, for instance, by $w_{ll}^2=w_{mm}^2=w_{0l}^2+w_{l0}^2=1$ does not exist as it would violate (\ref{ineq}). 
The derivation of (\ref{ABC3}) also proves that there is no other optimal state than (\ref{optimpro}).
Indeed, if the second and third term on the second line of (\ref{ABC3}) do not vanish identically, then the optimum is not reached. 
The optimality requires both $w_{ss}^2=0$ and $w_{ll}^2+w_{mm}^2=1$, and therefore $w_{ll}^2=1$ since $\ep_{ll}^2 \geq \ep_{mm}^2$.
The threshold   $\mu_\star$ is the value of $\mu$ for which $\ep_{ss}^2 + \ep_{mm}^2=2 \ep_l^2$. 
 With the notation $\delta_k\equiv1-\ep_k^2$, it reads
\BE
\label{musml}
\mu_\star=\frac{-\delta_m\ep_m^2-\delta_s\ep_s^2+\sqrt{2 \ep_l^2\left(\delta_m^2+\delta_s^2\right)-(\delta_m-\delta_s)^2}}{\delta_m^2+\delta_s^2}.
\EE

For $\mu_\star \leq \mu  \leq 1$, the ordering of the largest channel parameters is changed to $\ep_l^2 \leq \frac{1}{2}(\ep_{mm}^2+\ep_{ss}^2)$.
This yields 
\BEA
A+B+C &\leq&  2 \ep_l^2+\ep_{ll}^2+w_{ss}^2(\ep_{mm}^2+\ep_{ss}^2-2 \ep_{l}^2)\NN\\
&\leq&  \ep_{ll}^2 + \ep_{mm}^2 +\ep_{ss}^2.
\label{ABC4}
\EEA
The first line comes from the second one of (\ref{ABC3}) where the third term which is still negative or zero in the interval of $\mu$ considered here has been upper bounded by taking $w_{ll}^2+w_{mm}^2-w_{ss}^2=1$. On the other hand, the term $w_{ss}^2(\ep_{mm}^2+\ep_{ss}^2-2 \ep_{l}^2)$  is now positive and upper bounded by setting $w_{ss}=1$ which  gives the final result.
The  bound (\ref{ABC4}) is thus tight and achieved if and only if $w_{ss}^2=w_{mm}^2=w_{ll}^2=1$.
By (\ref{dec}), the optimal input states are the maximally entangled density matrices
\BE
\rho_{\Psi_\star}=\frac{1}{4} \left( \sigma_0 \otimes \sigma_0+\eta \, \sigma_1\otimes \sigma_1+\nu \, \sigma_2 \otimes \sigma_2+\xi \, \sigma_3 \otimes \sigma_3\right) ,
\label{optimbell}
\EE
which entails that $|\Psi_\star\rangle$ correspond to the Bell states  $\frac{1}{\sqrt{2}}(|00\rangle \pm |11\rangle)$ for $\pm  \eta=\mp \nu=\xi =1$ and   $\frac{1}{\sqrt{2}}(|01\rangle \pm |10\rangle)$ for $\pm \eta=\pm \nu= \xi =- 1$.
The eigenvalues of ${\mathcal E}(\rho_{\Psi_\star})$ are
\BEA
\lambda_{\eta,  \upsilon}=\frac{1}{4}(1 + \eta \ep_{33} +\upsilon [\ep_{11}+\eta\ep_{22}]) \qquad \eta, \upsilon=\pm 1 .\
\label{valBell}
\EEA

\emph {Illustration:} For  $q_0\!=\!0.2$, $q_1\!=\!0.1$, $q_2\!=\!0.3$, $q_3\!=\!0.4$,
the channel parameters (\ref{epn}) are $\ep_1\!=\!-0.4$, $\ep_2\!=\!0$, $\ep_3\!=\!0.2$, so that $l\!=\!1$, $m\!=\!3$ and $s\!=\!2$. Hence, up to $ \mu_{\star} \!\simeq \! 0.39$ the optimal states (\ref{optimpro}) are the product states  associated with eigenstates of $\sigma_1$. Notice that $\sigma_1$ is not the most probable transformation. This shows the relevance of the channel parameters: $\ep_1 \!\equiv \!q_0+q_1-q_2-q_3$ dominates because the rotations $\sigma_2$, $\sigma_3$  add up and are not compensated by $\sigma_0$, $\sigma_1$.

In conclusion, for two uses of arbitrary Pauli channels with memory modelled as a correlated noise, the amount of classical information which can be reliably transmitted per use is proven to be ${\mathcal C}_2(\mu)=1-\frac{1}{2}\sum_{\eta,\upsilon=\pm 1} \lambda_{\eta,  \upsilon} \log_2 \lambda_{\eta,  \upsilon}$ with $\lambda_{\eta,  \upsilon}\equiv \lambda_{\eta,  \upsilon}(\mu)$ given by (\ref{valpro}) for $0 \leq \mu  \leq \mu_\star$ and by (\ref{valBell}) for $ \mu_\star \leq  \mu  \leq 1$.   
Below $\mu_\star$, the capacity is achieved by  the tensor product of the
 single qubit density matrices pertaining to the eigenstates of the Pauli matrix $\sigma_l$ whose associated channel parameter $\ep_l$ has the largest absolute value.
Above the memory threshold,   the two-use classical capacity is reached by maximally entangled states.
Entanglement is thus a useful resource to enhance the transmission of classical information for this general class of quantum channels with memory.
The author is grateful to  N. J. Cerf and E. Karpov for simulating discussions.


\begin{thebibliography}{99}

\bibitem{H73} A.~S.~Holevo, IEEE Trans.~Inf.~Theory {\bf 44}, 269 (1998);
 B.~Schumacher and M.~D.~Westmoreland,  Phys.~Rev.~A {\bf 56}, 131 (1997).  
\bibitem{GLMSY04} V. Giovannetti, S. Guha, S. Lloyd, L. Maccone, J. H. Shapiro and H. P. Yuen, Phys. Rev. Lett. {\bf 92}, 027902 (2004).
\bibitem{BF05} A.~Sen(De), U.~Sen, B.~Gromek, D.~Bru\ss and M.~Lewenstein,
 Phys. Rev. Lett. {\bf 95}, 260503 (2005). 
\bibitem{A05} G.~G.~Amosov, Probl. Inf. Transm. {\bf 42}, 67 (2006).
\bibitem{Macch1} Ch. Macchiavello and G.~M.~Palma,  Phys. Rev. A {\bf 65}, 050301(R) (2002).
\bibitem{Macch2} Ch.~Macchiavello, G.~M.~Palma and S.~Virmani, Phys.~Rev.~A {\bf 69}, 010303(R) (2004).
\bibitem{BM05} G.~Bowen and S.~Mancini, Phys.~Rev.~A {\bf 69}, 012306 (2004).
\bibitem{BDM05} G.~Bowen, I. Devetak and S.~Mancini, Phys.~Rev.~A {\bf 71}, 034310 (2005).
\bibitem{KW05} D. Kretschmann and R. F. Werner, Phys. Rev. A {\bf 72}, 062323 (2005) 
\bibitem{GM05} V.~Giovannetti and S.~Mancini, Phys.~Rev.~A {\bf 71}, 062304 (2005).
\bibitem{CCMR05} N.~J.~Cerf, J.~Clavareau, Ch.~Macchiavello and J.~Roland, Phys.~Rev.~A {\bf 72}, 042330 (2005). 
\bibitem{RSGM05} G.~Ruggeri, G.~Soliani, V.~Giovannetti and S.~Mancini, Europhys. Lett. {\bf 70}, 719 (2005).
\bibitem{KDC06} E. Karpov, D. Daems and N. J. Cerf, Phys. Rev. A {\bf 74}, 032320 (2006).
\bibitem{AS} M. Abramowitz and A. Stegun, \emph{Handbook of mathematical functions}, (Dover, New York, 1972).
\end{thebibliography}
\end {document}